\documentclass{iaus}
\usepackage{graphicx}

\title[Summary Talk] 
{Symposium Summary}

\author[Claus Leitherer]   
{Claus Leitherer
}

\affiliation{Space Telescope Science Institute, 3700 San Martin Drive,
Baltimore, MD 21218, USA \\ email: {\tt leitherer@stsci.edu} }

\pubyear{2008}
\volume{250}  
\pagerange{1--10}
\jname{Massive Stars as Cosmic Engines}
\editors{F. Bresolin, P.A. Crowther \& J. Puls, eds.}
\begin{document}

\maketitle

\begin{abstract}
I summarize the highlights of the conference. First I provide a brief history of 
the beach symposia series our massive star community has been organizing. Then I
use most of my allocated space discussing what I believe are the main answered and open 
questions in the field. Finally I conclude with a perspective of the future of 
massive star research.

\keywords{sociology of astronomy, radiative transfer, telescopes, stars: early-type,
             ISM: evolution, (Galaxy:) open clusters and associations: general, galaxies: stellar content,
              (cosmology:) early universe, gamma rays: bursts}
\end{abstract}

\firstsection 
\section{Past}

This is the ninth meeting in what has become the traditional beach symposia series of the
massive star community. The series had its origins in 1968 at a workshop on Wolf-Rayet (W-R) stars in 
Boulder (definitely not a beach location). At that time, the cosmic significance of W-R
stars was largely unknown but understanding their nature was deemed important enough for
follow-up meetings. This then led to IAU Symp. 49 in Argentina in 1971, and to the 
subsequent tradition of holding massive star symposia at roughly five-year intervals. For
those who like statistics, here is the complete tally: 
IAU Symp. 83 (1978, Canada),
IAU Symp. 99 (1981, Mexico),
IAU Symp. 116 (1985, Greece),
IAU Symp. 143 (1990, Indonesia),
IAU Symp. 163 (1994, Italy),
IAU Symp. 193 (1998, Mexico),
IAU Symp. 212 (2002, Spain), and finally IAU Symp. 250 (2007, USA). 

Each meeting had its distinct flavor and theme. For instance, IAU Symp. 163 had the 
emphasis on binary stars, and IAU Symp. 116 allocated a large fraction of time to
massive stars in Local Group galaxies. Overall, a clear trend is apparent: the median
distance of the astronomical objects at each meeting has been steadily increasing. Distances were
expressed in kpc in the early symposia, then became Mpc, and during this meeting redshifts larger than
3 were commonly mentioned. 

Those who attended IAU Symp. 49 would have been astounded had they known how the field would have progressed by the time
of IAU Symp. 250. Even as recently as during IAU Symp. 193 (which I attended as well), it would have
been preposterous to discuss massive stars in connection with gamma-ray bursts, Lyman-break
galaxies, and Population III stars.

 I will finish my historical notes with a bit of trivia. Three
participants of this symposium have witnessed this rapid development from the very beginning.
They already participated in the 1971 meeting: Peter Conti, Lindsay Smith\footnote{I apologize
to Lindsay Smith whom I failed to name in my original talk.}, and Nolan Walborn. Peter Conti
distinguished himself with an impressive feat: he attended {\em all} nine beach symposia. His
record would have been tied were it not for Virpi Niemela's untimely death in 2006.

\section{Present}

The symposium {\em Massive Stars as Cosmic Engines} was structured along the five major science themes Atmospheres and Winds, Stellar Evolution, 
Nearby Populations, Feedback, and Early Universe. I will follow these themes and try 
highlighting the major issues that have emerged during the meeting.

\subsection{Atmospheres and Winds}

40 years after the construction of the first non-LTE model atmospheres and 30 years after the development of the 
radiatively driven wind theory the surfaces of OB stars are basically understood, and models and observations
agree quite well ({\em Puls}). Independent atmosphere codes used by different groups give consistent
results. We can routinely generate fully blanketed non-LTE models and compare them to observations. This 
has led to a major revision of the relation between spectral type and effective temperature ($T_{\rm{eff}}$),
the latter being 10 -- 20\% lower than previously thought. The ramifications for the ionizing photon output or the bolometric corrections are significant. The long-standing discrepancy between stellar spectroscopic and
evolutionary masses still persists, although the new atmospheres with their higher mass-luminosity ratio 
improve the agreement.

The dependence of the stellar mass-loss rate ($\dot{M}$) on heavy-element abundance ($Z$) is of particular
interest to those who need to extrapolate to metal-free Population~III stars. Empirical studies of hot
stars in the solar neighborhood, the Large (LMC) and the Small (SMC) Magellanic Cloud suggest a smooth
progress to lower rates in agreement with the wind theory ({\em de Koter}). The small lever in $Z$ makes
this conclusion a challenge: one would really like to perform such a study in truly metal-poor galaxies such
as, e.g., I~Zw~18 whose oxygen abundance is 1/20 that of the LMC. 

Of course, the devil is in the detail. While we understand the overall physics and derived parameters of OB stars,
the effects of micro- and macroturbulence, variability, porosity, and wind clumping are still poorly understood
({\em Hillier}). Wind clumping, if not taken into account properly, leads to an overestimate of $\dot{M}$. Back in
the 1980s and 1990s we were overly confident in our ability to determine mass-loss rates. It has become clear
that many of the published values (including those by myself) need to be revised downward.

Wind instabilities can be traced with thermal X-ray emission ({\em Cohen}). Interestingly, two classes of
X-ray emitters are found among (single) massive stars: those with soft and with hard keV spectra. The former are
associated with shocks in the wind whereas the latter are related to magnetic fields. Soft X-ray spectra are found
in evolved O stars. In contrast, the hard X-ray emitters tend to be located closer to the zero-age main-sequence.
The caveat is the small sample size of the observed stars, which precludes studying dependencies other than on age. 

It was gratifying to learn that the atmospheric modeling of W-R stars has reached the state-of-the-art of that
for O stars ({\em Crowther}). We believe we understand the W-R parameters as well (or as badly) as those of O stars.
This is quite an improvement in comparison with the 1980's when W-R luminosities were uncertain by an order of
magnitude. One of the key points we should all take away from this meeting is {\em WN}~$\approx O \neq$~{\em WC}.
Most WN's (in particular those of the latest types) are core-hydrogen burning O stars in disguise. They may still be
close to the main-sequence. WC stars are truly devoid of hydrogen.

Empirical hydrodynamical models of W-R winds allow one to determine mass and mass-loss-rate ({\em Gr\"afener}). This
opens up the possibility of using $\dot{M}$ as a proxy for the W-R mass. In cases when the most massive stars in
a populous star cluster, like the Arches cluster in the Galactic Center, are dominated by WN-like objects, one can
utilize the W-R mass-loss rates to probe the initial mass function (IMF).

Luminous Blue Variables (LBVs) have made a strong come-back at this meeting. LBVs were a hot commodity in
the 1980s and 1990s, when they were thought to be an essential phase in the evolution of a massive star. However,
they took the backseat -- quite undeservedly -- at the last two symposia. Now they are back in full force. One
reason is the revived interest in an evolutionary phase with strong mass loss. Since $\dot{M}$ in all pre- and 
post-LBV phases has been recognized to be lower than assumed, an additional source for removing stellar material
is needed. Needless to say that LBVs are fascinating objects which deserve to be studied in their own right.

The physical mechanism responsible for the giant LBV outbursts differs from radiation pressure. The key is 
the location of LBVs close to the Eddington limit. In this case, the mechanical luminosity of a continuum
driving wind can in principle reach the radiative luminosity limit ({\em Owocki}). The Eddington limit may
even be temporarily exceeded when pulsations are taken into account ({\em Onifer}). Such pulsations are triggered
by a build-up of luminosity when the interior reaches a temperature of $\sim$200,000~K, corresponding to the
Fe opacity peak. A special case may be $\eta$~Car. The fact that the star is a binary may help understand 
some of the properties of the circumstellar material ({\em Okazaki}). The audience was divided as to whether
binarity can explain the entire LBV phenomenon. 

The cool side of the uppermost Hertzsprung-Russell (HR) diagram often looks rather un-cool to the massive star community: 
red supergiants (RSGs) are rarely covered in this symposium series. I was glad to learn about a fresh analysis
of an RSG sample with MARCS atmospheres that result in substantially higher $T_{\rm{eff}}$ and lower luminosity ($L$).
When placed on the HR diagram, the RSGs are now in excellent agreement with stellar evolution models ({\em Massey}).
The  most luminous RSGs are surrounded by resolved material. They offer one of the few occasions when stellar
astronomers can directly observe the 2-dimensional structure of a star, rather than inferring it from point 
source spectroscopy. The morphology of the circumstellar material is direct evidence of stellar magnetic fields and
photospheric convection ({\em Humphreys}).

The royal method of determining stellar masses utilizes binary systems. Binaries hosting the most luminous, and 
therefore the most massive stars are of particular interest. There are only a few of them but their analysis uniformly 
suggests a mass of about 100~$M_\odot$ for the most massive stars ({\em Moffat}). This is consistent with the
upper mass limit found for the IMF in rich clusters. Interestingly, mid/late WN stars turn out to be the most massive stars known.
All of them are core-hydrogen burning.

\subsection{Stellar Evolution}

The stellar evolution modelers delivered another message we all should take home: {\em stellar rotation is a
fundamental channel leading to the formation of W-R stars} ({\em Maeder}). Originally, all W-R stars were suggested to
be components of binaries, with orbital forces being responsible for the mass loss. In 1976, Peter Conti
recognized that stellar winds in single stars could remove mass as efficiently, and a 2nd W-R channel was 
established. 30 years later, rotation becomes the 3rd channel. At low $Z$, rotation may even be the dominant
formation channel ({\em Meynet}).

Many model predictions of stellar evolution must rely on sometimes rather daring extrapolations. This is particularly true when it comes to evolution to very low $Z$. I was impressed about the diligence of the modelers who try to test
their predictions using all angles of observational evidence: not only do they perform the classical HR diagram comparisons, but they also seek
guidance from the WN/WC ratio, the nitrogen surface abundance, or the pulsar rotation velocities. The
fact that so vastly different constraints lead to consistent answers gives confidence in the basic soundness of the 
models.

We can even dare to make predictions for stellar evolution at essentially zero metal abundance ({\em Hirschi}). What matters is 
rotation, rotation, rotation.... Centrifugal forces lead to stronger mechanical and radiatively driven winds, the
latter being enhanced via chemically enhanced material that is brought to the surface. Since even the first stellar
generations effectively switch from the pp to the CN cycle and synthesize metals, one may expect significant mass loss
already early on.

Stellar pulsations are ubiquitous across the HR diagram but evolution models generally ignore them. Exploratory
calculations suggest that pulsations do in fact modify rotation and trigger additional mixing ({\em Townsend}). Perhaps the
buzz word at the next beach symposium will be pulsation, pulsation, pulsation....

Binary evolution can play a crucial role for understanding the observed properties of individual stars and of massive
populations as a whole. Mass transfer affects evolution in many ways, one of them being increased spin-up and therefore
even higher rotation velocity. Could all rapid rotators be binary stars ({\em Langer})? Population models accounting
for binary evolution appear to improve the agreement between predicted and observed stellar number ratios 
({\em Eldridge}). Regrettably, the stellar census in Local Group galaxies is still incomplete and/or subject to large
statistical errors because small number statistics. Clearly, a push for more extensive surveys is needed.
 
Most (all?) massive stars end their lives as core-collapse supernovae (SNe). It has been a long-standing challenge
to revive the shock during the core-collapse ({\em Burrows}). Traditional models need an ad hoc mechanism to prevent
the shock from stalling, such as an artificial ``piston'' or ``bomb''. The key to making the core-collapse self-consistent
is the geometry: once the explosion is assumed to be non-spherical, instabilities will set in, and the shock will
not stall.

Being affiliated with STScI, I was delighted to see the good use of the Hubble archive in the search for SN
progenitors. Many potential host galaxies of core-collapse SNe have been observed with one of the HST imagers. Chances
are, the SN progenitor has been captured by HST and its properties can be extracted from the HST archive ({\em Smartt}).
The typical SN II progenitor is a RSG with a mass of about 10~$M_\odot$. This is the progenitor canonically expected
for a type II SN, suggesting that SN1987a's blue progenitor really was an oddball confusing the issue. SN2006gy
seems to be yet another oddball. Its properties hint at a dense circumstellar environment ({\em N. Smith}). One
possible explanation is an LBV progenitor. Could this occur more commonly than previously thought?

Magnetars are neutron stars with ``insanely'' high magnetic fields of order $10^{15}$~G. A variety of independent
arguments suggest an association with very massive progenitors ({\em Gaensler}). We now believe that 10\% of all
massive stars end their lives as magnetars. While this is not the dominant end phase, it is certainly not
insignificant, either. Extremely massive stars are predicted to evolve into pair-instability SNe. Since Population~III
stars are expected to be strongly biased towards very high masses, there should be many such pair-instability
SNe in the early universe. The absence of any evidence in favor of their existence (e.g., from abundance patterns)
may indicate that pair-instability SNe do not form in large numbers. How can we avoid their formation ({\em Ekstr\"om})?
Rotation again leads the way: primordial massive stars may not enter the pair-instability SN phase if they suffer from
rotation induced mass loss and have strong magnetic fields.

Gamma-ray bursts (GRBs) were one of the major themes of this conference. There is now convincing evidence that long
GRBs are located in star-forming galaxies and are associated with SNe of type Ibc ({\em MacFadyan}). Lifetime arguments favor
a small star with a hydrogen-free envelope. This then leads to a progenitor model with a massive, rotating star. Both
single and binary models are being discussed for GRB progenitors ({\em Yoon}). The observed diversity of GRBs may be
accounted for by magnetic fields with a range of properties. It
is truly remarkable to witness the progress we have made on this subject since the last massive star symposium. 

The host galaxies of nearby long GRBs can be studied in some detail. The chemical composition is of particular
interest. GRBs are preferentially found in relatively small, metal-poor galaxies ({\em Stanek}). The GRB hosts,
when plotted on a metallicity vs. luminosity diagram, tend to fall below the mean relation. The implication is that GRBs
trace only low-metallicity star formation. Obviously, life on Earth benefits from this bias. It is unlikely that a
long GRB will be found in our own Milky Way, something we should not be too keen about.

\subsection{Nearby Populations}

We witnessed a major change of guard in the session on massive-star populations. 30~Doradus used to be the gold 
standard and Rosetta Stone, yet we heard very little about this region. Rather, the heavily dust-obscured star clusters in
our own Galaxy attracted a lot of the attention during this meeting. There are currently 10 massive ($\sim$10$^5 M_\odot$) star
clusters known in our Galaxy, although this is just the tip of the iceberg ({\em Figer}). Notable examples are
the Arches and Quintuplet clusters near the Galactic Center. The Arches cluster is dense and rich enough to allow 
a full sampling of the IMF well in excess of 100~$M_\odot$ if such massive stars existed. However, no stars more massive than about 150~$M_\odot$ are found. This is interpreted as a genuine upper limit to the IMF.

The Galactic Center is the birth site of 30\% of the W-R stars known in our Galaxy. Many of these W-R stars are still
hydrogen-rich and should actually be considered O stars ({\em Martins}). The W-R stars near the Galactic Center serve
as a warning for those using the presence of such stars as a cluster age indicator. Finding such ``pseudo-W-R'' stars
does not necessarily indicate an evolved cluster age. The corresponding ages are well below 2~Myr. The same caveat applies to the
well-known Galactic cluster NGC~3603.

The large reddening towards the Galactic Center requires the infrared (IR) spectral region for atmospheric analyses. Non-LTE models are now available that allow us to determine abundances of both $\alpha$- and Fe-group elements ({\em Najarro}). The abundances found in hot and cool supergiants as well as in the surrounding ionized gas are consistent and suggest roughly solar composition. These techniques are highly relevant for current and upcoming observing facilities which emphasize the near-IR window.

Westerlund~1 is the most massive young star cluster known in our Galaxy. Will it become the new Rosetta Stone of
star clusters? Since Westerlund~1 is heavily dust obscured, the IR is the wavelength of choice. The cluster has a mass
of order $10^5 M_\odot$ and an age of 4.5~Myr ({\em Negueruela}).  There is a significant binary population as 
indicated by numerous discrete X-ray sources. The most pressing issue for resolution is the uncertain distance. Like
most Galactic clusters we do not know the distance to better than about 20\%, which is a serious limitation in comparison with extragalactic clusters or the clusters near the Galactic Center.

Westerlund~1 should have many brethren in the Galaxy. Where are they? Extrapolating a standard cluster luminosity function for our Galaxy leads to an expected number of about 100 such clusters ({\em Hanson}). Most of them would have been missed in previous surveys. New searches are underway to complete the census.

Ultracompact HII (UCHII) regions are at the lower end of the HII luminosity function. The recently completed GLIMPSE survey with the Spitzer Space Telescope turns out to be a treasure trove for identifying UCHII regions ({\em Conti}). UCHII regions are in a
pre-Orion evolutionary state with ages of a few $10^5$~yr. Their ionizing stellar population is tiny by Westerlund~1 standards. There are typically only just a few ionizing stars. An important result is the trend to find no isolated ionizing stars. They all come in small clusters, which is consistent with the suggestion that most (all?) massive stars are not born in isolation.

If massive stars live most of their lives in clusters, it becomes important to address the interplay between stellar evolution and the dynamical evolution of the cluster ({\em Vanbeveren}). Close stellar encounters between single and
multiple stars affect both stellar and cluster evolution. Some stars will merge and/or be ejected from the cluster. As a result, runaway stars would be the products of stellar mergers.  

Several recently completed surveys of nearby star-forming galaxies with, e.g., GALEX and Spitzer have provided us with a panchromatic view of extragalactic massive star populations ({\em Bresolin}). Comparison of IR and ultraviolet (UV) luminosities suggests there is no major hidden star formation in {\em normal} galaxies. The stellar light is attenuated by about $A_{\rm V} = 1$ but otherwise no significant part of the stellar population is missed in the UV/optical. We can now measure stellar and nebular abundances in a variety of nearby galaxies. The comparison indicates agreement to within a factor of 2.

We were reminded that star formation can occur in unexpected environments, such as in the dSp galaxy Fornax and in the dE galaxy NGC~205 ({\em Grebel}). Significant star formation occurred in both galaxies as recently as $\sim$100~Myr ago. Asymptotic giant branch (AGB) stars are the sign posts of that star-formation episode. While AGB stars are not massive stars by most definitions, they are close to RSGs in many of their properties. In particular, AGB stars and RSGs have similar stellar wind physics since their winds are both thought to be dust driven. 

{\em Spectroscopists do it better (Kudritzki)}. ``Extragalactic stellar astrophysics'' was another buzzword at this meeting. Owing to new instrumentation {\em and because of the vision and drive of dedicated astronomers} we are able to obtain stellar
spectra in galaxies beyond the Local Group whose quality rivals those obtained in the Galaxy. Large surveys like FLAMES have been mentioned numerous times at this meeting. Its impact cannot be overemphasized. Targeted observations of individual stars out to a distance of about 7~Mpc via spectroscopy provides us with reddening independent distances. We were told that $H_0$ may not be as well known as we thought...

Some 25 years ago Phil Massey and Tony Moffat and their groups pioneered the systematic search for W-R stars in Local Group galaxies. In 2007, such surveys can be done in NGC~1313 at a distance of 4.1~Mpc ({\em Hadfield}). Pushing beyond the Local Group is highly complementary to the more detailed studies in, e.g., M33 or M31. We sacrifice individuality for completeness.

\subsection{Feedback}

The term ``feedback'' kept appearing throughout this symposium. This is one theme which crosses the boundaries of traditional hot star research and embraces topics such as the interstellar medium (ISM) dynamics, galaxy evolution, and the first stellar generations. 

We observe nebular structures around stars on scales ranging from $10^1 - 10^3$~pc and with ages of $10^4 - 10^7$~yr ({\em Chu}). Depending on their sizes and origination, these structures are commonly referred to as bubbles, superbubbles, and supergiant shells. 30 years ago, Weaver et al. published their classic paper describing the evolution of a circumstellar bubble with simple analytical scaling relations. These relations are still commonly used. Unfortunately, real bubbles tend not to follow the relations very well. Only when the detailed radiation-hydrodynamics are taken into account, modeled and observed bubbles agree ({\em Arthur}). In particular, observations clearly show complex substructure, like clumps, whose presence turns out to be crucial for the modeling.

HD~316285 is an extreme P~Cygni star which has been proposed as an LBV candidate. A recent Spitzer mid-IR survey of circumstellar matter around Galactic star detected circumstellar emission around this star. Surprisingly, the distance to the mid-IR nebula turns out to be $\sim$8~kpc, much farther than the canonical (uncertain) distance to HD~316285. If correct, this distance would lift HD~316285 into the status of the most luminous star known in our Galaxy, essentially identical to the Pistol star. Since the uncertainties are still large, we may not see a press release anytime soon. 

Feedback is fundamental at any redshift and at any time. We were given beautiful examples of the regulatory forces of feedback at redshifts 30, 2 and 0 ({\em Dopita}). Molecular hydrogen is the major coolant on the primordial ISM. Star formation sets in once the gas has cooled down sufficiently. When the newly formed stars inject wind and SN material, the H$_2$ may be removed, thus effectively curbing the cooling and star formation process. Redshift 2 is the main epoch of massive galaxy formation. These galaxies host massive central black holes which can trigger galaxy-wide outflows. The outflows remove material from the galaxy centers, inhibiting further growth. Finally, local starbursts inject matter and energy into the ISM and may eventually pollute the surrounding intergalactic medium (IGM).

M82 is for the starburst community what $\zeta$~Pup, $\eta$~Car, or 30~Dor are for those working on individual hot stars. Its numerous super star cluster exemplify the very nature of a starburst: {\em it is the intensity that makes a starburst (Gallagher)}. Super star clusters have masses of order $10^6 M_\odot$ and form rapidly over $10^6$~yr. The corresponding star-formation rate of $1 M_\odot$~yr$^{-1}$ is comparable to that of the Milky Way, with the notable difference that the super star cluster has a size of $\sim$1~pc. The term ``socialized photoionization'' was introduced, meaning that starburst regions in a galaxy do not exist in isolation but may mutually share their photon output and affect each other's evolution.    
 
The dwarf galaxy NGC~1569 is one of the closest starbursts and consequently offers a detailed view of its ISM morphology ({\em Linda Smith}). Spectroscopy of the bottom of the outflow of NGC~1569 leads to a picture where the outflow consists of a superposition of individual superbubbles. There is no evidence yet for a highly organized galactic superwind as observed farther out in X-rays.

What is the ionization source of the diffuse ISM ({\em Oey})? This long-standing open question has been brought back on the agenda owing to the revised relation between stellar spectral type and ionizing photon output (see above). Previously, there seemed to be an excess of photons in HII regions that could leak out and heat the diffuse ISM. The new calibration essentially eliminates this excess, and the ionizing source of the diffuse ISM is again an open issue. 

Massive star yields are often considered as being determined by Type II supernova yields alone. While it is true that SNe are often the dominant nucleosynthetic source, some subtle but important observations can only be understood when stellar wind yields are taken into account ({\em Matteucci}). For instance, we observe an excess of nitrogen at very low chemical composition. This excess is most readily understood as due to primary nitrogen produced in massive stars. Another example is [O/Mg], which plotted vs. O/H is not constant. Intuitively (at least to my intuition), one would expect the ratio of two $\alpha$-elements to be constant. However, pre-SN stellar winds in massive stars modify the oxygen abundance prior to the SN explosion.

Despite a general understanding of the core-collapse SN explosion mechanism, the predicted yields for heavy
elements like $^{45}$Sc, $^{49}$Ti, and $^{64}$Zn are lower than indicated by observations of metal-poor stars ({\em Fr\"ohlich}). These (and other elements) are produced in the core of the SN where the details of the explosion mechanism matter. The fact that we know that a supernova explodes in principle, is not sufficient for computing the corresponding yields!

\subsection{Early Universe}

The conference organizers devoted a full session to the high-redshift universe --- the largest allocation in the history of the beach symposium series. Spectroscopic features of massive stars in high-$z$ galaxies (the so-called Lyman-break galaxies) were first reported in the mid-1990s. By the time of IAU Symp. 212 the quality and quantity of the spectra had reached a state to permit IMF studies using the strongest lines. Now, we have reached the stage when faint absorption features are used to study abundances ({\em Pettini}). Star-forming galaxies at $z \approx 3$ have oxygen abundances already close to solar. This means galaxies observed at a lookback time of about 10~Gyr have almost the same metal content as galaxies in the local universe.  

Lyman-$\alpha$ emitters are another class of star-forming galaxies at cosmological redshift ({\em Taniguchi}). They are the lower luminosity extension of Lyman-break galaxies. Lyman-$\alpha$ emitters can be found at higher redshift and lower luminosity because a few percent of the entire starburst power is concentrated just in the Lyman-$\alpha$ line. These galaxies are now routinely found out to redshift 7. The Subaru telescope has turned out to be one of the most prolific discovery machines. Galaxies harboring genuine Population~III stars are predicted to have a strong nebular He~II 1640 line. Searches for this line are underway, so far with no firm detections. Since the prediction entirely hinges on the correct radiative transfer prediction in the stellar atmosphere below 228~\AA, one would definitely like to perform some independent local tests of the models before allocating too many resources to such searches. 

The early universe shows ample evidence of dust. For instance, the spectra of quasars at high redshift imply enormous dust masses. What is the source of this dust? Core-collapse SNe are likely candidates, and Spitzer was used to search for dust features around SNe in the mid-IR ({\em Kotak}). Surprisingly little dust was detected leading to the speculation that a significant amount of dust may be hidden in clumps.

GRB host galaxies have by now been detected out to the highest redshifts. Thanks to the rapid response of the Swift satellite, the host galaxies of tens of GRBs have been identified, and many of them could be studied in great detail ({\em Fynbo}). The hosts tend to be blue and irregular but display an enormous diversity in parameters such as reddening, Lyman continuum escape, or hydrogen column density. So far, most of the information on the hosts refers to the galaxy as a whole. Obviously one would like to make the direct association between the GRB and the progenitor and its birth place. One route that is being followed is the analysis of damped Lyman-$\alpha$ absorbers in GRB afterglow spectra ({\em Dessauges-Zavadasky}). However, the spectra analyzed so far do not exhibit circumstellar material but are dominated by the general galaxy ISM. The direct signatures of GRB progenitors are still elusive.

Hypernovae are a particular subclass of SNe whose kinetic energies are an order of magnitude or more above the canonical value of $10^{51}$~erg ({\em Nomoto}). There is fairly good evidence that hypernovae originate from progenitors with masses more massive than about 40~$M_\odot$. Morphologically they are associated with SNe of type Ic. Several long-duration GRBs have been firmly linked to these objects. The nucleosynthesis in hypernovae differs from that of standard supernovae by generating elevated carbon and diminished iron yields. This trend may in fact by observed in very metal-poor stars, emphasizing the importance of hypernovae in the early universe.  

There is a fairly strong conviction among theorists that the first generation of stars was born with a top-heavy IMF ({\em Bromm}). The absence of dust and a higher equilibrium temperature favor higher accretion during the star-formation process. As a result, the characteristic masses of stars are an order of magnitude higher than in the present-day universe, and the masses of the most massive stars may be as high as 500~$M_\odot$. This is a research area that is begging for close
collaborations between star formation experts, atmosphere modelers and cosmologists: such extremely massive and hot stars are ideal testbeds for the predictions of, e.g., the ionizing radiation in the extreme UV and the mass loss at zero metallicity. 

We heard about a rather subtle imprint of the first stellar generation in the cosmic IR background ({\em Kashlinsky}). In principle, existing Spitzer mid-IR spectra carry the information of this early star formation. However, extracting the signal and separating it from other effects is a formidable task, and final verification may have to await the James Webb Space Telescope.

\section{Future}

I am sure all participants will agree with me that this symposium was a resounding success. The field of massive stars is healthy and shaping many other ``hot'' topics in contemporary astronomy. For everyone's entertainment, I am presenting in Table~\ref{ADS} the result of a little exercise prior to the symposium. I was asking myself these questions: What is the impact of massive stars? How did the impact evolve over time? How does it compare to other fields of astronomy? My approach was very simplistic and would not withstand a truly rigorous analysis. Nevertheless I hope one can recognize a few trends that some of you may find useful.  

The entries in Table~\ref{ADS} were generated as follows: I searched the ADS for the keywords in the first column, binned in five-year intervals from 1975 until 2005. The last column of the table gives the ratio of the entries in the 2000--2005 over the 1975--1980 columns. For instance, ``galaxies'' was found in 6812 articles between 1975 and 1980, and 41,087 times between 2000 and 2005. The increase is the result of more publications with time, but also the increasing completeness of the ADS, large number of preprints, etc. Therefore the numbers should be looked at only in a relative sense.

The first three entries are the generic keywords ``galaxies'', ``interstellar'', and ``stars''. Their frequency increases by factors of a few, with galaxies showing the steepest rise. These numbers are the benchmarks for comparison. The next seven entries are related to stellar astronomy. My reading of these numbers is as follows. Keywords traditionally associated with hot, massive stars are essentially flat (i.e., just mirroring stellar astronomy). This does not suggest all stellar subcategories are flat: ``brown'' and ``AGB'' show a significant increase.

The next six entries are typically associated with extragalactic publications. They display a clear upward trend, even neglecting the 1975--1980 column where statistics are small. Note, for instance the increase of ``survey'' in the past 15 years, reflecting publications related to the major extragalactic surveys.

The divergent trends of ``ultraviolet'' and ``infrared'' are quite illuminating. They trace the increasing numbers of IR facilities and related results and the relative decline of the field of UV astronomy. The latter has traditionally been more closely connected with massive stars than the former. The final two entries compare ``Hertzsprung-Russell'' and ``color-magnitude''. This again signals a shift towards color-magnitude diagrams using photometric techniques with wide-field detectors as opposed to HR diagrams constructed from spectroscopy of individual stars.   

Here is my reading of the lessons learned from this little exercise.
\begin{itemize}
\item{The number of papers dealing with ``classical'' topics related to massive stars is lagging behind some other subjects in stellar astrophysics, such as AGB stars. The hot star community should be more aggressive in leveraging their scientific expertise and take ownership of fields where stellar astrophysics of massive stars adds value. The subject of Population~III stars in the early universe is an excellent example. We saw this happening at this meeting but stronger efforts are needed.}
\item{On average, the growth of many extragalactic/cosmological topics exceeds that of stellar topics. While this may seem disconcerting to some of us, this should not necessarily be the case. The separation between stellar and extragalactic is becoming increasingly fuzzy and meaningless. For instance, ``starbursts'' are more and more becoming embraced by the massive star community. At the same time, the extragalactic community is studying local template stars for comparison with Lyman-break galaxies.} 
\item{The growing emphasis on the IR is clear and present. This community should take advantage of this wavelength regime in areas where progress is lacking: understanding the formation of massive stars, dissecting heavily obscured regions like the Galactic Center, and developing diagnostic techniques to utilize spectral information at these wavelengths. These efforts will pay off when the next generation of large grand-based and space telescopes will be functional.}
\item{I am speculating that one of the reasons for the waning popularity of the HR diagram may be the trend towards surveys and wide-field observation techniques. If so, we may see a revival of spectroscopically determined stellar properties with the advent of several multi-object/integral field unit spectrographs over the next few years. The results of the FLAMES project are an example of the enormous scientific returns to be expected.}
\end{itemize} 

\begin{table}
  \begin{center}
  \caption{Popularity of some keywords over time as returned by ADS.}
  \label{ADS}
  \begin{tabular}{l | c c c c c c c}\hline 
{\bf Keyword} & {\bf 75--80} & {\bf 80--85} & {\bf 85--90} & {\bf 90--95} & {\bf 95--00} & {\bf 00--05} & {\bf (00--05)/(75--80)} \\  \hline

galaxies &	6812&	9928&	13556&	17649&	32400&	41087&	6.0\\
interstellar&	6458&	8091&	9433&	10017&	9337&	10234&	1.6\\
stars&	16862&	23227&	26788&	29264&	42483&	52792&	3.1\\
     \hline						
AGB&	1&	82&	344&	883&	2100&	2796&	2796\\
atmosphere&	7949&	8228&	8278&	8871&	9641&	14621&	1.8\\
brown&	121&	135&	326&	707&	1742&	3000&	24.8\\
non-LTE&	179&	217&	355&	435&	644&	765&	4.3\\
OB&	369&	631&	618&	747&	1075&	1826&	4.9\\
supernova&	2099&	2747&	5221&	5203&	7890&	12735&	6.1\\
Wolf-Rayet&	297&	760&	869&	952&	1135&	1120&	3.8\\
	\hline					
AGN&	1&	103&	632&	1706&	3723&	6514&	6514\\
bulge&	219&	408& 658&	1215&	2297&	2944&	13.4\\
redshift&	1527&	2170&	3090&	5000&	11066&	16010&	10.5\\
reionization&	3&	3&	17&	75&	287&	1049&	350\\
starburst&	1&	97&	737&	1454&	3236&	4702&	4702\\
survey&	1752&	2812&	3870&	6501&	12631&	22836&	13.0\\
	\hline					
ultraviolet&	3808&	5759&	5419&	6141&	6520&	7473&	2.0\\
infrared&	4872&	6439&	9987&	11770&	16975&	23487&	4.8\\
\hline
Hertzsprung-Russell&	370&	683&	547&	435&	237&	282&	0.8\\
color-magnitude&	144&	344&	730&	916&	1164&	1326&	9.2\\ \hline
  \end{tabular}
 
 \end{center}
\end{table}

The field of massive stars will greatly benefit from the new generation of large ground-based telescopes ({\em D'Odorico}) and JWST ({\em Sonneborn}). I am excited to learn about the results at future beach symposia of the massive star community.

\end{document}